\documentclass[preprint]{aastex}
\usepackage{amsmath,amssymb}
\usepackage{bmpsize}
\usepackage{graphicx}
\usepackage{natbib}
\usepackage{color}
\usepackage[colorlinks,citecolor=blue,linkcolor=red]{hyperref}
\usepackage{longtable}
\usepackage{threeparttable}
\usepackage[figuresright]{rotating}
\usepackage{tablefootnote}

\def\neowise{{\it NEOWISE}}

\begin{document}

\title{An On-going Mid-infrared Outburst in the White Dwarf 0145+234: 
Catching in Action of Tidal Disruption of an Exoasteroid?} 
\author{Ting-gui Wang\altaffilmark{1,2}, Ning Jiang\altaffilmark{1,2}, 
Jian Ge\altaffilmark{3}, Roc M. Cutri\altaffilmark{4}, Peng Jiang\altaffilmark{5}, 
Zhengfeng Sheng\altaffilmark{1,2}, Hongyan Zhou\altaffilmark{1,5},
James Bauer\altaffilmark{6}, Amy Mainzer\altaffilmark{7}, 
Edward L. Wright\altaffilmark{8} } 
\altaffiltext{1}{CAS Key Laboratory for Researches in Galaxies and Cosmology,
University of Sciences and Technology of China, Hefei, Anhui 230026, China; 
twang@ustc.edu.cn}
\altaffiltext{2}{School of Astronomy and Space Sciences, University of Science and Technology of China, Hefei, 230026, China}
\altaffiltext{3}{211 Bryant Space Science Center Department of Astronomy University of Florida Gainesville, FL 32611-2055}
\altaffiltext{4}{IPAC/Caltech, 1200 E. California Blvd., Pasadena, CA 91125 USA}
\altaffiltext{5}{Antarctic Astronomy Research Division, Key Laboratory for Polar Science of the State Oceanic Administration, Polar Research Institute of China, Shanghai, China}
\altaffiltext{6}{Department of Astronomy, University of Maryland, College Park, MD 20742, USA}
\altaffiltext{7}{University of Arizona, 1629 E University Blvd Tucson AZ 85721,
USA}
\altaffiltext{8}{Division of Astronomy and Astrophysics, University of California at Los Angeles, CA  90095, USA}
\begin{abstract}
We report the detection of a large amplitude MIR outburst in the white 
dwarf (WD) 0145+234 in the NEOWISE Survey data. The source had a stable MIR 
flux before 2018, and was brightened by about 1.0 magnitude in the W1 and W2 
bands within half a year and has been continuously brightening 
since then. 
No significant variations are found in the optical photometry data during the 
same period. This suggests that this MIR outburst is caused by recent 
replenishing or redistribution of dust, rather than intrinsic variations of the WD. 
SED modeling of 0145+234 suggests that there was already a dust disk 
around the WD in the quiescent state, and both of the temperature and surface 
area of the disk evolved rapidly since the outburst. 
The dust temperature was $\simeq$1770K in the initial rising phase, close to the 
sublimation temperature of silicate grains, and gradually cooled down to around 1150K,
while the surface area increased by a factor of about 6 during the same period. 
The inferred closest distance of dust to the WD is within the tidal 
disruption radius of a gravitationally bounded asteroid. We estimated the 
dust mass to be between $3\times 10^{15}$ to $3\times 10^{17}$ $\rho/(1 \mathrm g~cm^{-3})$ kg 
for silicate grains of a power-law size distribution with a high cutoff size from 0.1 
to 1000$\mu$m. We interpret this as a possible tidal breakup of an exo-asteroid 
by the WD. Further follow-up observations of this rare event may provide insights 
on the origin of dust disk and metal pollution in some white dwarfs. 
\end{abstract}


\section{Introduction}

A substantial fraction of white dwarfs (WDs) are known to show metal absorption 
lines in their spectra (Zuckerman et al. 2010; Koester et al. 2014). Because 
heavy elements diffuse out of the photosphere in a rather short time (days to 
hundred years, depending on $T_{eff}$ and log$g$) in the strong surface 
gravity of a WD \footnote{http://montrealwhitedwarfdatabase.org/evolution.html} (Koester 2009; Fontaine et al. 2015),
this indicates that these heavy elements were added to the surface very recently, 
presumably, by accretion of tidally disrupted asteroids or comets 
(Jura \& Young  2014). Many WDs also display infrared excesses, which may be 
considered as evidence for dust disks (Zuckerman,\& Becklin 1987; Farihi, 
Jura,\& Zuckerman 2009; Rebassa-Mansergas et al. 2019). The presence of a 
cold disk is further supported by the detection of double-peaked infrared 
Ca II emission lines (Jura 2003;G\"änsicke et al. 2006; Veras et al. 2014). 
Both dust/gas disks in WDs are either transient or continuously replenished with 
 new dust because grains can be either expelled from the disk through 
radiation pressure for small grains or dragged onto the star through the 
Poynting-Robertion force on time scales of several years (Rafikov 2011).  
Fragmentation of a planetary body has been proposed as a source of dust (Jura 
\& Young 2014; Vanderburg et al. 2015; Manser et al. 2019). Variability in 
infrared flux on time scale of less than a year has been reported very 
recently in a number of WDs (Xu et al 2018; Swan, Farihi, \& Wilson 2019), 
but most of variations are generally of small amplitudes (e.g., 20-40\%, 
Xu \& Jura 2014; Xu et al. 2018; Swan et al. 2019) on time scales 
of a year. 

In this letter, we report discovery of a large MIR outburst in the white 
dwarf WD 0145+234 (01:47:54.81 +23:39:43.6, Wills \& Wills 1974). This outburst 
has so far lasted one and a half year. This event 
was discovered during a blind search for large amplitude MIR outbursts in 
the WISE/NEOWISE data archive (Wright et al. 2010; Mainzer et al. 2014). 
This could be a process of tidal disruption caught in action. 

\section{Data and Analysis}

\begin{figure*}
\figurenum{1}
       \centering
        \begin{minipage}{0.85\textwidth}
        \centerline{\includegraphics[width=1.0\textwidth]{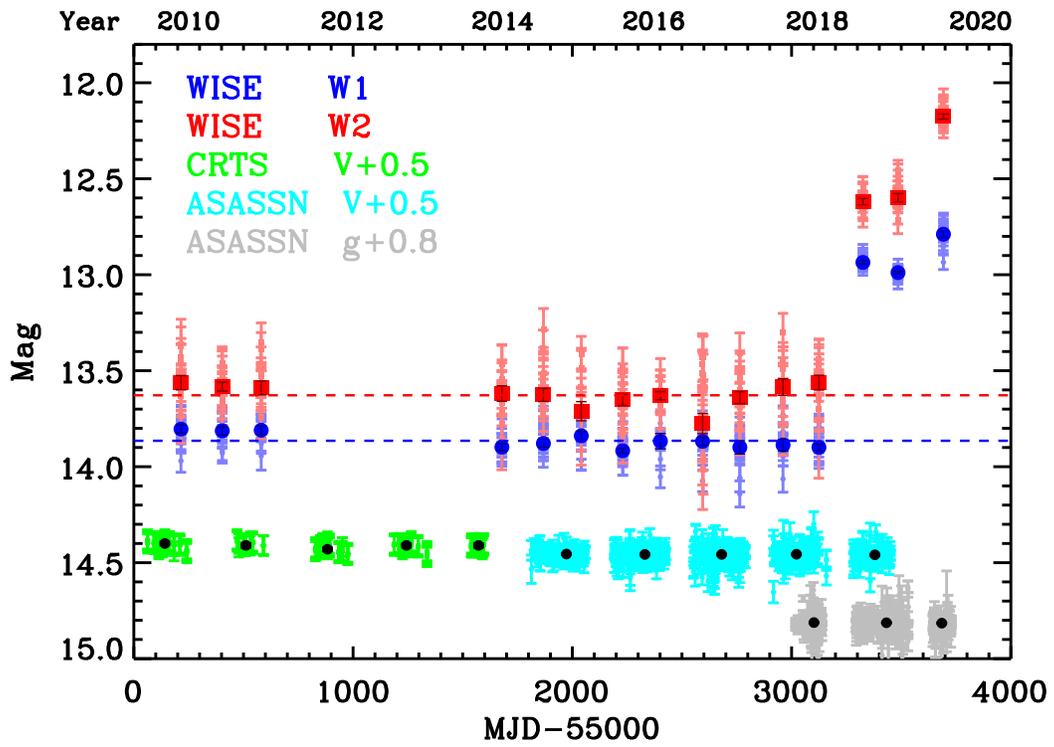}}
        \end{minipage}
        \caption{The WISE/NEOWISE(-R) and optical light curves of the white dwarf 0145+234. 
Legend: blue circles and red squares are the epoch median of W1 and W2 magnitude. Black 
circles are median magnitudes in V or g over one observational season.    
\label{fig:lcs}}
\end{figure*}

The WISE all-sky survey and NEOWISE Reactivation mission conducted repeated scans of the entire sky at 3.4 and 4.6 microns (hereafter W1 and W2) beginning in January 2010 through the present, except for the period February 2011 to November 2013.
We retrieved the W1 and W2 point spread function (PSF) profile-fit 
photometry of 0145+234 from the AllWISE Multi-epoch Photometry Table 
and \neowise-R 
Single Exposure (L1b) Source Table~\footnote{https://irsa.ipac.caltech.edu/cgi-bin/Gator/nph-scan?mission=irsa\&submit=Select\&projshort=WISE},
which contain all measurements from 2010 to July 2019.
The single-exposure data were first screened by the quality flag 
marked in the catalogs to remove measurements with poor quality or 
possible corruption: ($qual\_frame < 5$), charged particle hits ($saa\_sep < 5$), 
scattered moon light ($moon\_masked=1$), and artifacts ($cc\_flags>0$).
The high quality measurements in each six month observation epoch were averaged to 
increase
the signal-to-noise ratio (S/N) of the photometry, resulting in 15 epochs 
(see Figure~\ref{fig:lcs}).
No significant variations are present in the first 12 epochs 
and thus we take them as the quiescent state with average magnitudes of 
$13.87\pm0.04$ and $13.63\pm0.06$ in the W1 and W2 bands, respectively.
The light curve displays a large outburst between the January 2018 and July 2018 
epochs ($\Delta W1=0.93\pm0.04$, $\Delta W2=1.01\pm0.06$) and continues 
rising in the later epochs. 
Overall, the outburst appears to display a redder-when-brighter trend in which 
the variation amplitude in the W2 band is larger than that in the W1 band.

The brightening of 0145+234 is not the result of object’s motion carrying it close to a bright nearby source. The proper motion of the white dwarf is measured by Gaia to be $\mu_{RA}= -5.21\pm0.12$ and $\mu_{DEC} = -97.59\pm0.08$ mas/year (Gaia Collaboration 2018).  Examination of the AllWISE Atlas Image (epoch 2010) shows that the closest MIR source is approximately 14 arcsec to the southwest of the white dwarf position.  The nearby source was cleanly separated and measured in the AllWISE Catalog, and is 1.6 magnitudes fainter in W1 and W2 than J0145+234 in 2010.  Even if the white dwarf moved directly onto that source, which it did not (see Figures S2 and S3), the apparent flux increase would be much too small to account for the brightening observed in the light curve.

The relative motion of 0145+234 is clearly detected from the AllWISE and NEOWISE astrometry that spans a time baseline of 9 years and is fully consistent with proper motion measured by Gaia, as illustrated in Figure S1.  The position of the white dwarf during outburst does not show any deviation from the expected motion larger than 0.1 arcsec.  Thus, a chance coincidence with a bright nearby source could not have shifted the photocenter of the white dwarf, making this occurrence unlikely.

W1 and W2 images comparing the region around 0145+234 in January 2014, January 2018 and January 2019 are shown in Figures S2 and S3.  The January 2014 and 2018 images are pre-outburst, and the difference images between the 2018 and 2014 show only a very small positive/negative residual due to the slight motion of the source in the four years separating the observations.  The differences between the 2019 and 2014 images show a bright, point-like image at the position of the white dwarf that is the source in outburst.

0145+234 is bright in the optical band ($V\sim$13.93 mag) and thus has been 
well measured by various optical time-domain surveys.
We retrieved the optical photometry data from public released data archive from  
the Catalina Real-Time Transient Survey~\footnote{http://nunuku.caltech.edu/cgi-bin/getcssconedb\_release\_img.cgi} (CRTS; Drake et al. 2009) 
and All-Sky Automated Survey for Supernovae (ASAS-SN, Shappee et al. 2014; Kochanek et al. 2017)~\footnote{https://asas-sn.osu.edu/}.
The CRTS survey has monitored 0145+234 since 2005 without filters, but the photometry is 
calibrated to a pseudo-$V$ magnitude using a few dozen pre-selected standard stars in each field.
The public CRTS data are available to October 2013.
Fortunately, the public ASAS-SN survey can serve the subsequent $V$-band photometry. 
Although the latest observations are performed in the $g$-band, these $g$-band data 
largely overlap with V-band data. In contrast to the remarkable MIR variability, 
0145+234 is quite stable and shows negligible variability in the long-term (more 
than one decade) optical light curves, including the MIR outburst period.

We made the quiescent SED of the WD by collecting data from GALEX, PANSTARRs 
(Chambers et al. 2016), Gaia (Gaia Collaboration et al. 2016), 2MASS (Skrutskie et al. 2006) 
and ALLWISE. We matched the UV to near-infrared photometry with 
the synthesized WD SED models 
\footnote{http://www.astro.umontreal.ca/~bergeron/CoolingModels}. 
The models cover the range of $T_{eff}$ from 2500 K to 90,000 K and 
log $g$ from 7.0 to 9.0 for DA WDs, and $T_{eff}$ from 3250 K to 70,000 K 
and log$g$ from 7.0 to 9.0 for DB WDs (Tremblay, Bergeron, \& Gianninas 2011; 
Bergeron et al. 2011; Blouin et al. 2018).
The interstellar reddening of the CCM-law with $R_V=3.1$ (Cardelli, Clayton, \& Mathis 1989) was added 
with the E(B-V) as a free parameter. The best fitted parameters are listed in Table 1.  
They are consistent with those derived from the spectroscopic model (T=13060$\pm$217 K, 
log$g$=8.13$\pm$0.05; Gianninas 2011). With these photospheric parameters, we also derived 
other parameters of the WD \footnote{http://montrealwhitedwarfdatabase.org/evolution.html}
(Gianninas et al. 2011; Fontaine et al. 2001): the mass of $M_*=0.667\,M_\sun$, the radius of $R_{WD}=0.0116$ 
$R_\sun$, the luminosity of $L_{WD}$=0.00350 $L_\sun$, and the age of 0.381Gyr. 

\subsection{IR excess in the low state}

The observed ALLWISE fluxes in the W1, W2 and W3 (12$\mu$m) bands are clearly above the predicted values from the WD model. 
The infrared excess indicates a dust disk. Initially, we added a black-body curve to 
model the infrared excess, and found that it is insufficient to fit all the data with a 
2$\sigma$-excess in the W3 band, which requires an additional temperature component 
or the non-grey dust model to fit. However, since a two-component model needs a total of 4 
parameters (two temperatures and two surface areas), while only three data point (W1, 
W2 and W3) are available, such a model cannot be fully constrained. To illustrate 
the possibility of non-grey dust model, we adopt the Particle-Cluster-Aggregation (PCA) 
model for olivine (Nakamura 1998), which is used to explain the scattering disk of 
$\beta$ Pic. Without increasing the number of free parameters, the model can now fit the data. 

\subsection{Dust in the high state }

In the high state, this source is more than one magnitude brighter in the W1 and 
W2 bands. Since we only have two data points, we chose to fit the MIR excess with 
a single temperature black body model. We considered two different scenarios: 1) the 
quiescent dust disk has been transformed into the latter one; 2) 
the quiescent disk remains the same and there is a new dust component that 
contributes to the brightening of the source. In the first case, we fit the total observed 
excess in the W1 and W2 band at high state with a black body model.  In the second 
case, we only fit the varying W1 and W2 fluxes with a black-body model to constrain the 
newly formed dust component. The varying fluxes were calculated by subtraction of the 
mean quiescent flux from the high state fluxes, and uncertainties were calculated through 
propagations of errors. These results are summarized in table \ref{table:bbfit}.  In the first 
case, the temperature was increased from 1163K to 1769 K from quiescent epochs (1-12) to 
epoch 13 while the surface area might be increased but the large uncertainties in the 
quiescent state make this inconclusive. 
 In the subsequent epochs, the dust temperature 
is decreased continuously to 1146 K at the last epoch,  while the surface area increases 
by a factor of about 4. In the second case, the newly formed dust had initial temperature about 
1950K, and decreased to 1115K at the last epoch, while surface area increased by a factor of 
about 6.

\begin{table}
\centering\scriptsize
\caption{Black body model fit to the MIR excess\label{table:bbfit}}
\begin{tabular}{ccc ccc ccc}\\
\hline
\hline
epoch   &  W1  & W2  &\multicolumn{3}{c}{scenario 1\tablefootnote{fitting excesses to the WD model}} & \multicolumn{3}{c}{scenario 2\tablefootnote{fitting quiescent state subtracted MIR flux} } \\
        &      &     &  T           &  A           & R             &  T  &  A   & R   \\
        & mag  & mag &  K           & 10$^{16}$~m$^2$        & $R_\sun$      &  K  & 10$^{16}$~m$^2$        & $R_\sun$  \\\hline
 1-12  & 13.87$\pm$0.04 & 13.63$\pm$0.06 & 1163$\pm$145 & 3.54$\pm$1.50& 1.46$\pm$0.37  & \nodata    & \nodata &\nodata      \\
 13       & 12.94$\pm$0.01 & 12.62$\pm$0.02 & 1769$\pm$74 & 4.80$\pm$0.51& 0.63$\pm$0.05 &1949$\pm$529& 2.99$\pm$1.88 & 0.52$\pm$0.28 \\
 14       & 12.99$\pm$0.01 & 12.60$\pm$0.02 & 1515$\pm$62 & 6.96$\pm$0.84& 0.88$\pm$0.07 &1570$\pm$329&4.89$\pm$2.77 & 0.80$\pm$0.34 \\
 15       & 12.79$\pm$0.02 & 12.18$\pm$0.01 & 1146$\pm$32 & 23.06$\pm$2.08& 1.52$\pm$0.09 &1115$\pm$130&21.34$\pm$ 8.80& 1.59$\pm$0.37 \\
\hline
\end{tabular}
\end{table}

With the above fit parameters, we can estimate physical quantities of the dust. The distance to the 
WD is 29.458$\pm$0.044 pc using the parallax from the Gaia DR2 data (Gaia Collaboration et al. 
2016; 2018; Lindegren et al. 2018), which is consistent with that derived from the SED 
fitting. We adopted the bolometric luminosity of 0.0035$L_\sun$ and WD radius of 0.0116 
$R_\sun$ from the spectroscopic WD model \footnote{http://montrealwhitedwarfdatabase.org/evolution.html} 
(Gianninas et al. 2011; Fontaine et al. 2001). Assuming the dust is grey and in 
thermal equilibrium, we estimated the distance to the central star is on the order of
\begin{equation}
R_{dust} = \left(\frac{L_{WD}}{4\pi\sigma T_d^4}\right)^{1/4}=\left(T_{eff}/T_d\right)^2R_{WD}.
\end{equation}
The distances of the dust to the WD are also listed in Table 1. The closest distance to the WD is about 0.5 $R_\sun$.
 
The covering factor of the dust can be estimated from the ratio of the black-body luminosity 
to the WD luminosity assuming dust is optically thick to the UV to optical light of the WD. 
The covering factor in the high state sharply rises to around $(1.5-2.0)\times 10^{-2}$ from 
$(2.7\pm0.3)\times 10^{-3}$ in the quiescent state, and remains nearly constant in the last three 
epochs.  
  
Real dust is likely not grey, so the derived temperature and dust mass depends on the 
size distribution of grains ($N(a)$) and their absorption coefficients ($Q$). 
In the following, we assume that the sizes of dust grains follow a power-law distribution 
with an index $n$ in the range from $a_{min}$ to $a_{max}$ and none outside the range. 
The dust emission flux can be estimated
\begin{align}
f_\nu &=\int_{a_{min}}^{a_{max}} N(a) 4\pi a^2 Q_\nu(a) da \pi B_\nu(T)/(4\pi d_L^2)\nonumber \\
     &=\frac{3M_d}{4\rho d_L^2}\left<\frac{Q_\nu}{a}\right> B_\nu(T) \\
\end{align} 
where 
\begin{equation}
\left<\frac{Q_\nu}{a}\right>=\frac{\int_{a_{min}}^{a_{max}}N(a) a^3(Q/a) da}{\int_{a_{min}}^{a_{max}}N(a) a^3 da};
\end{equation}
and $M_d$, $\rho$ are the mass and density of the dust, respectively. The $d_L$ is the distance to the WD.
In the black body case, the absorption coefficient is $Q_\nu=1$. 

Using silicate coefficients\footnote{For grains of size larger than 10$\mu$m, we set  
$Q_\lambda\simeq 1$ for radiation at wavelengths greater than 4 microns.} in Laor \& Draine 
(1993), we calculated a set of $\left<\frac{Q_\nu}{a}\right>$ for $a_{min}$=0.01$\mu$m and 
$a_{max}=$0.1,\ 1,\ 10,\ 100,\ 1000$\mu$m. The excess infrared emission is then fitted with 
each of above models to derive mass and temperature of the dust. The lowest 
derived-temperatures are $T_d$=770, 1031, 948, 800K for epoch 1-12, 13, 14, 15, respectively, 
or 60-70\% of those from the black body model, when $n=-4$ and $a_{max}=0.1\,\mu$m. The highest 
temperatures are within 99\% of those from the black-body model when $n=-1$ and 
$a_{max}=1000\,\mu$m. 
The mass of dust is in the range of $3\times 10^{15}\,\rho_1$ to 3$\times$10$^{17}\rho_1$ kg 
($\rho_1=\rho/(1$ g~cm$^{-3}$)), depending strongly on the size distribution. A strict lower limit 
on the dust mass is a few times $10^{15}$ kg, obtained when most grains have a size of about 1$\mu$m, 
which gives the maximum ${Q_\nu}/{a}$ in the W1 and W2 band. For a given distribution, dust mass 
increases monotonically with time. The increase of the amount of dust can be produced by further 
break-up of large bodies due to collision. 
 
\begin{figure*}
\figurenum{2}
   \centering
   \begin{minipage}{0.85\textwidth}
   \centering{\includegraphics[angle=90,width=1.0\textwidth]{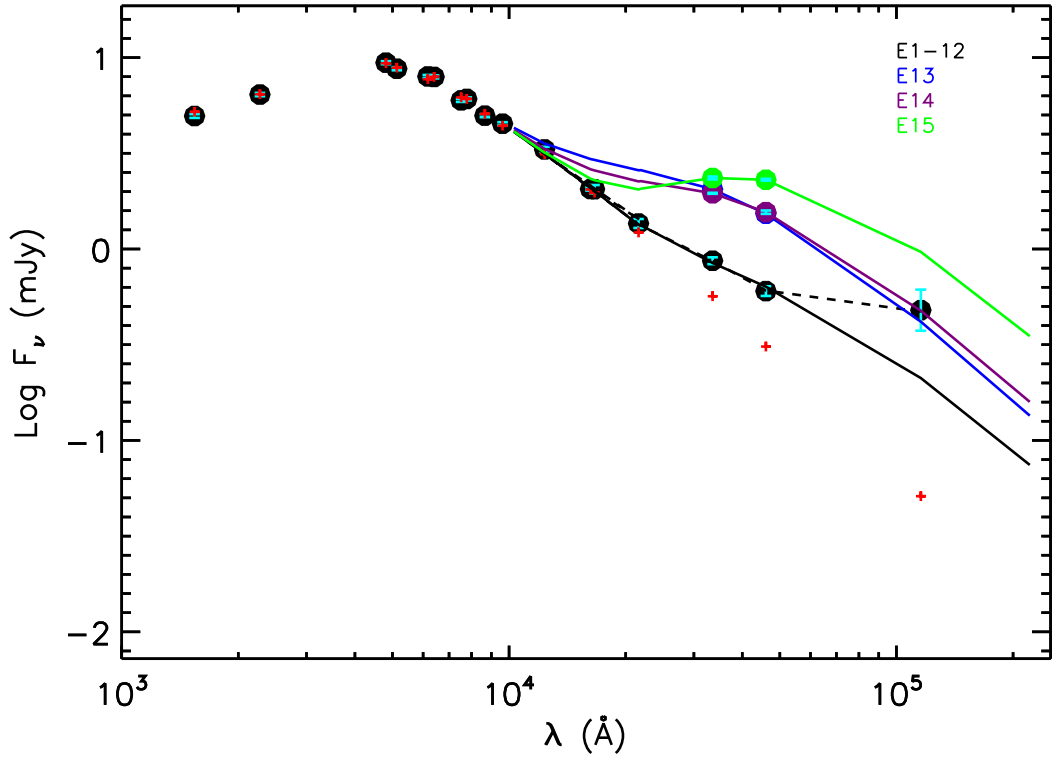}}
   \end{minipage}
   \caption{The spectral energy distribution (circles with error bars in cyan) and best fitted models 
(cross or solid curve). The black, blue, purple and green symbols represent the observed data on 
epoch 1-12, 13, 14 and 15 in the light curve, respectively (see table \ref{table:bbfit}). The WD 
photosphere model is represented by red crosses. The solid curves represent the black body fit 
to the excesses over the WD photosphere model (colors follow the symbols). The PCA dust model is shown as the dashed green curve.     
\label{fig:sedfit}}
\end{figure*}

\begin{figure*}
\figurenum{3}
   \centering
   \begin{minipage}{0.85\textwidth}
   \centering{\includegraphics[angle=90,width=1.0\textwidth]{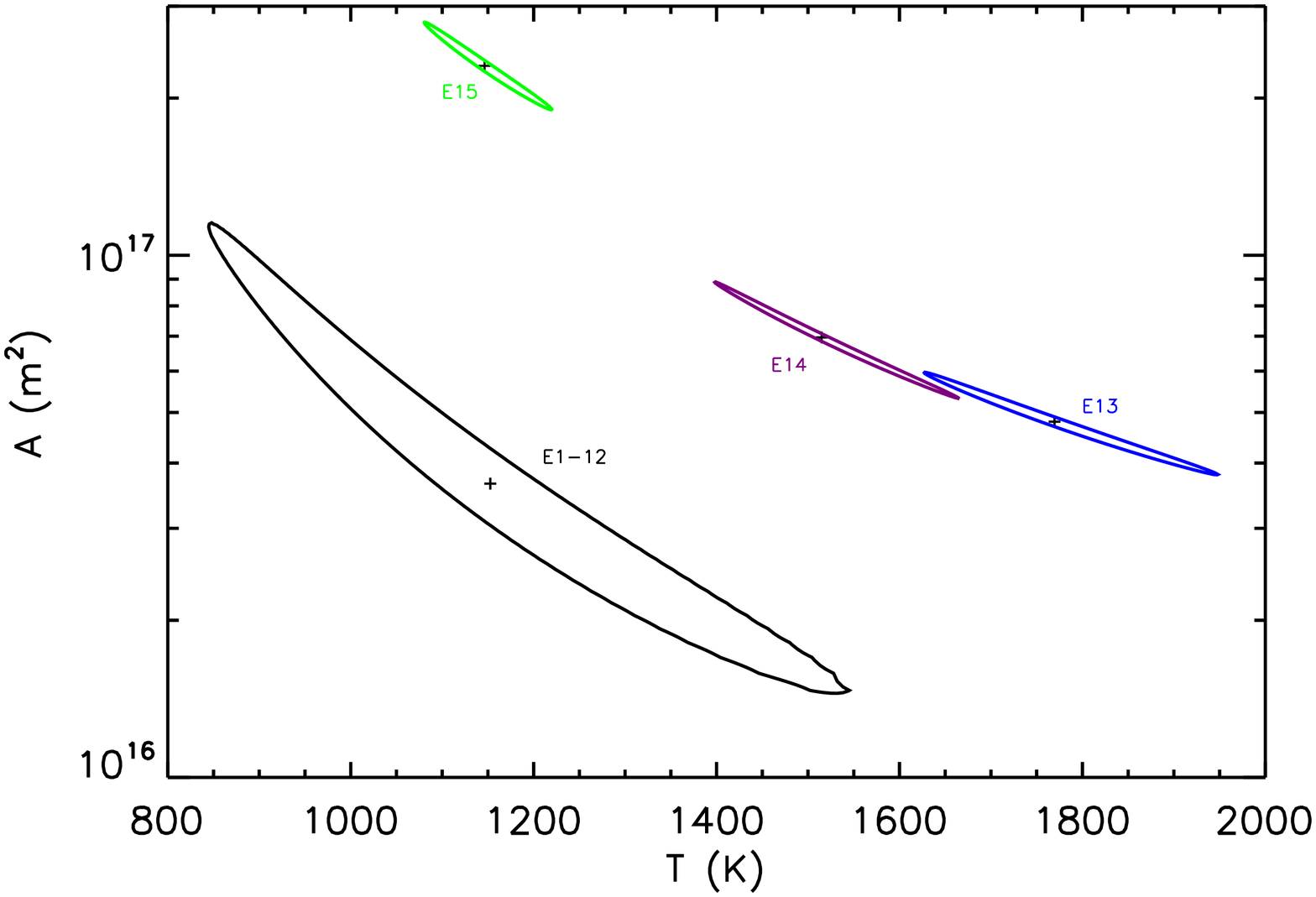}}
   \centering{\includegraphics[angle=90,width=1.0\textwidth]{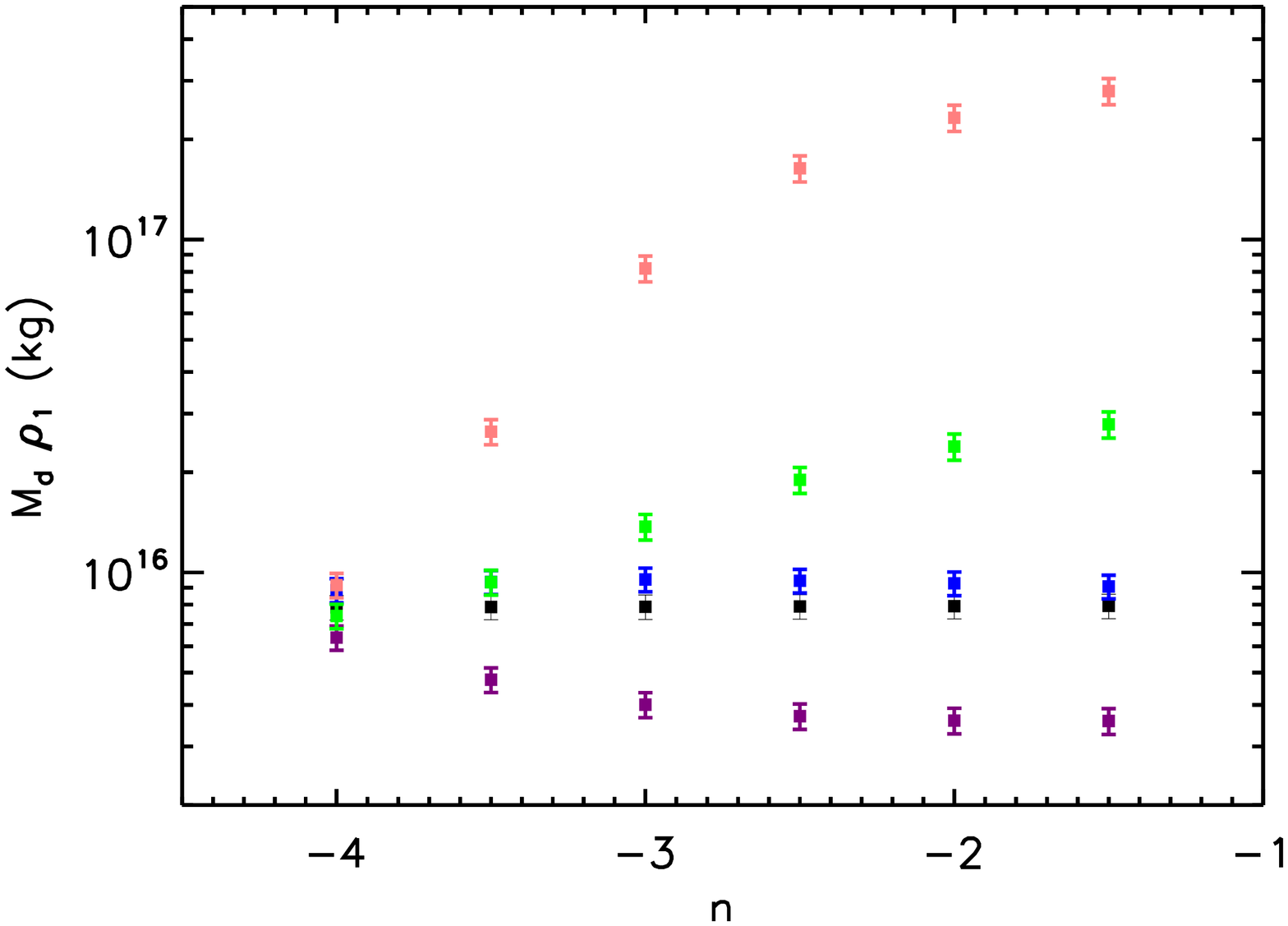}}
   \end{minipage}
   \caption{Upper panel: 90\% contours for the blackbody fit to the excess IR flux on four 
epochs. The best fitted parameters are marked as crosses. Colors are the same as those in Figure 
\ref{fig:sedfit}. Lower panel: the fitted dust mass of silicate grains as a function of power-law 
index at different cutoff of the largest grain size (red:1000, green:100, blue: 10, black:1, 
and purple: 0.1 $\mu$m) for the last epoch of the NEOWISE-R survey. 
\label{fig:contour}}
\end{figure*}

\section{Discussion}

The infrared light curve consists of a stable pre-burst quiescent phase, a short 
time rising outburst within half a year, and the MIR excess increasing phase with little 
variations in the optical light curve during the same period. This suggests that there 
is either a rapid inflow of an initial dust disk seen in the quiescent state into the 
inner region or tidal break-up of an asteroid into a clump of grains. 

In order to understand various processes which may contribute to the dust changes, 
we examined the time scales of various processes, for the distance of dust to the host 
star of $\sim$1 $R_\sun$, the WD mass of 0.667$M_\sun$ ,and the luminosity of 0.0035$L_\sun$. 
First, the dynamic time, or the orbit period, is about half an hour. 
Because we do not have the detailed light curve through the rising phase, 
we cannot reject the scenario that the newly 
formed dust is being brought by a giant comet. However, the late cooling (on time 
scale of years) is too slow for a departing comet. It is highly unlikely that the event was 
caused by a passing comet. 
   
Grains spiral onto the host star due to radiation drag (the Poynting-Robertson drag) 
on a time scale of (Burns, Lamy \& Soter, 1979; Backman \& Paresce 1993)
\begin{equation}
t_{PR}=4.73\left(\frac{b}{1\mu m}\right) \left(\frac{\rho}{g~cm^{-3}}\right)^2\left(\frac{R}{R_\sun}\right)^2\left(\frac{0.0035L_\sun}{L_{WD}}\right)\frac{1}{1+albedo} {\rm yr},
\end{equation}
i.e.,a few years in this case. Destruction of a large grain into small ones through collisions takes 
place on a time scale of (Backman \& Paresce 1993) 
\begin{equation}
t_{col}=3.5 \left(\frac{R}{R_\sun}\right)^{3/2}\left(\frac{0.667M_\sun}{M_{WD}}\right)^{1/2}\left(\frac{10^{-5}L_\sun}{L_{IR}}\right) {\rm yr},
\end{equation}
which is long, $\sim$ 10 years for the quiescent dust disk. Accretion of dust due to friction 
in the dust disk is much longer than $10^4$ years (Rafikov 2011; Girven et al. 2012).  Given that this outburst event time scale 
appears to be consistent with the radiation drag and collision dust destruction time scales, it is most likely that the dust is 
produced by tidal break-up of an asteroid-like object. 

The MIR observations are consistent with this picture. The temperature in the 
initial rising phase is as high as 1770K in the black-body fit. Assuming the dust is in 
radiative equilibrium, we estimate the distance of the dust to the WD is about $(0.63\pm 0.05)$ 
$R_\sun$, which is consistent within a tidal disruption radius, $r_{t}= 0.69\left
(\frac{M_{WD}}{0.667M_\sun}\frac{3 {\rm g~cm}^{-3}}
{\rho}\right)^{1/3}R_\sun$ of a gravitationally bounded asteroid with a typical density of 
$\rho=(1-7)$ g~cm$^{-3}$ (Carry 2012). The disrupted debris will have a range of specific 
energies and be spread in space. A small fraction with the lowest energies may fall 
directly onto the WD, and pollute the surface of the WD. A large part of debris may form an 
eccentric disk. Collision in the clump of debris will further cause fragmentation
, releasing smaller grains, and thus increases the surface area of dust as observed. 
Similar events were detected before, for instance, evidence for 
planetesimals in an orbit close to the tidal disruption radius was reported in WD 1145+017 
through the transit signal (Vanderburg et al. 2015; Xu et al. 2016). More recently, 
Manser et al. (2019) reported detection of a planetesimal orbiting within the debris 
disc through its perturbation to the gas disk. Our discovery may be the first case of an asteroid 
break-up process caught in action. 

\section{Conclusion and Perspective}

We report a large MIR outburst of the white dwarf WD 0145+234, which is still rising in the WISE  
W2-band. We interpret this event as tidal disruption of an asteroid by the WD. As the 
source MIR flux is still rising, further monitoring of this event will likely help provide 
insights to the fate of planet systems at their end of star lives. High resolution 
spectroscopic observations in the optical and UV can trace composition of recent falling debris (asteroid, e.g.,Swan et al. 2019). 
According to the calculation given by Fontaine\footnote{http://montrealwhitedwarfdatabase.org/evolution.html}, the diffusion time for 
most metals are less than a month.
The disruption may also release volatile gas into the interplanetary space, leaving 
absorption lines in the UV and optical spectrum (Wilson et al. 2019). The 
debris may contain warm gas emitting infrared CaII emission lines,  which can be used to trace 
the kinematic motion of the debris. Infrared photometric observations at long wavelengths and MIR 
spectroscopy can be also used to constrain dust composition.        

This work is supported by Chinese Science Foundation (NSFC-11833007,11421303).   
This research makes use of data products from the Wide-field Infrared Survey Explorer,
which is a joint project of the University of California, Los
Angeles, and the Jet Propulsion Laboratory/California Institute
of Technology, funded by the National Aeronautics
and Space Administration. This research also makes use of
data products from NEOWISE-R, which is a project of the
Jet Propulsion Laboratory/California Institute of Technology,
funded by the Planetary Science Division of the National
Aeronautics and Space Administration. This research has made use of the NASA/ IPAC Infrared Science Archive, 
which is funded by the National Aeronautics and Space Administration and operated by the California Institute of Technology.

\begin{figure*}
\figurenum{S1}
   \centering
   \begin{minipage}{0.95\textwidth}
   \centering{\includegraphics[angle=90,width=1.0\textwidth]{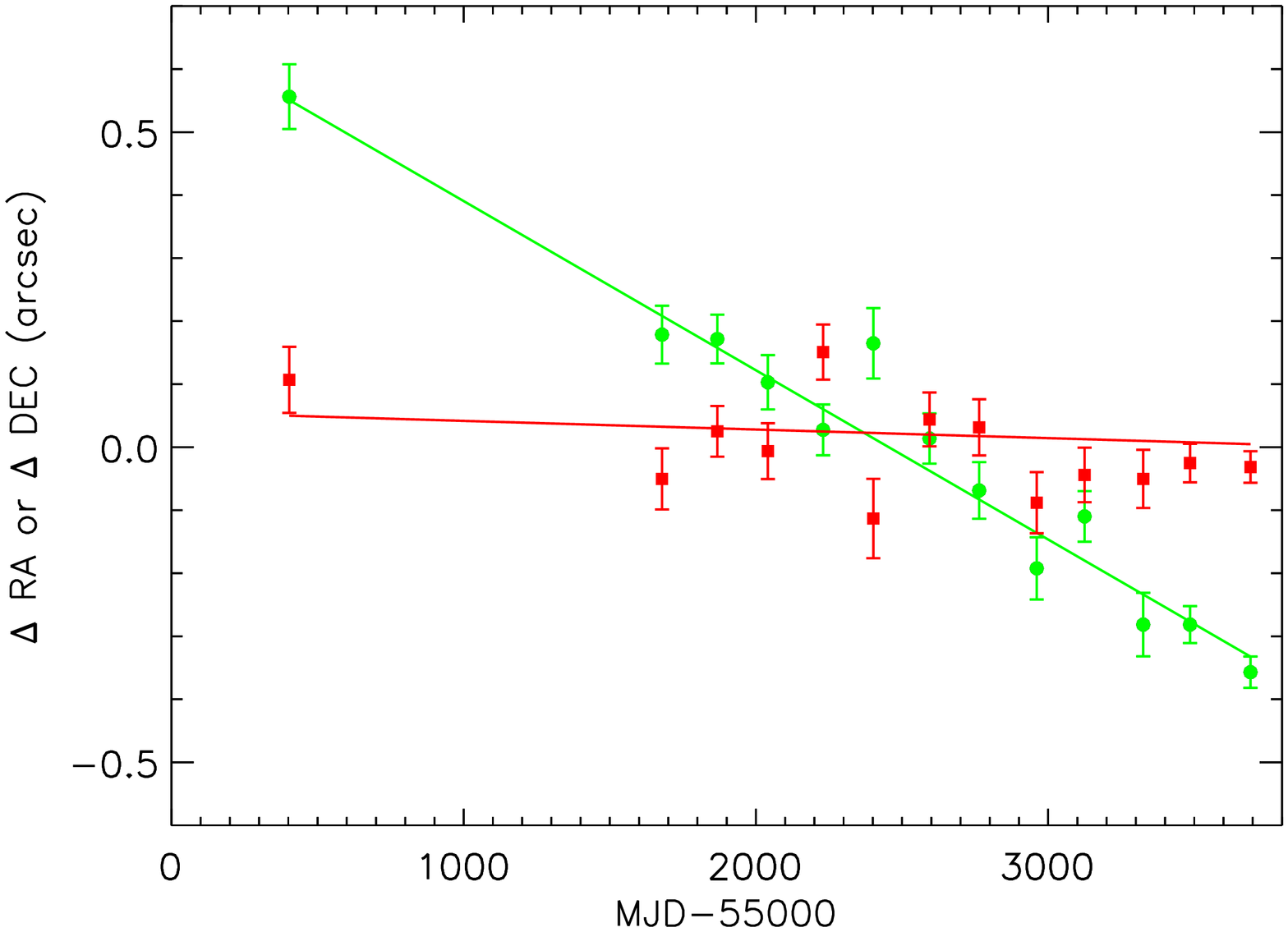}}
   \end{minipage}
   \caption{Relative RA (in red) and Dec (in green) motion of the MIR source with respect to their mean values during the WISE/NEOWISE surveys.  The earliest points are from the AllWISE Catalog and represent the average positions from the Juanuary 2010, July 2010 and January 2011 measurement epochs.  The points following MJD=55600 are the average positions measured during each six month NEOWISE observation epoch.  For comparison, the solid red and green lines show the expected RA and Dec motion of the white dwarf based on the Gaia astrometry and proper motion.
\label{fig:position}}
\end{figure*}

\begin{figure*}
\figurenum{S2}
   \centering
   \begin{minipage}{0.99\textwidth}
   \centering{\includegraphics[width=1.0\textwidth]{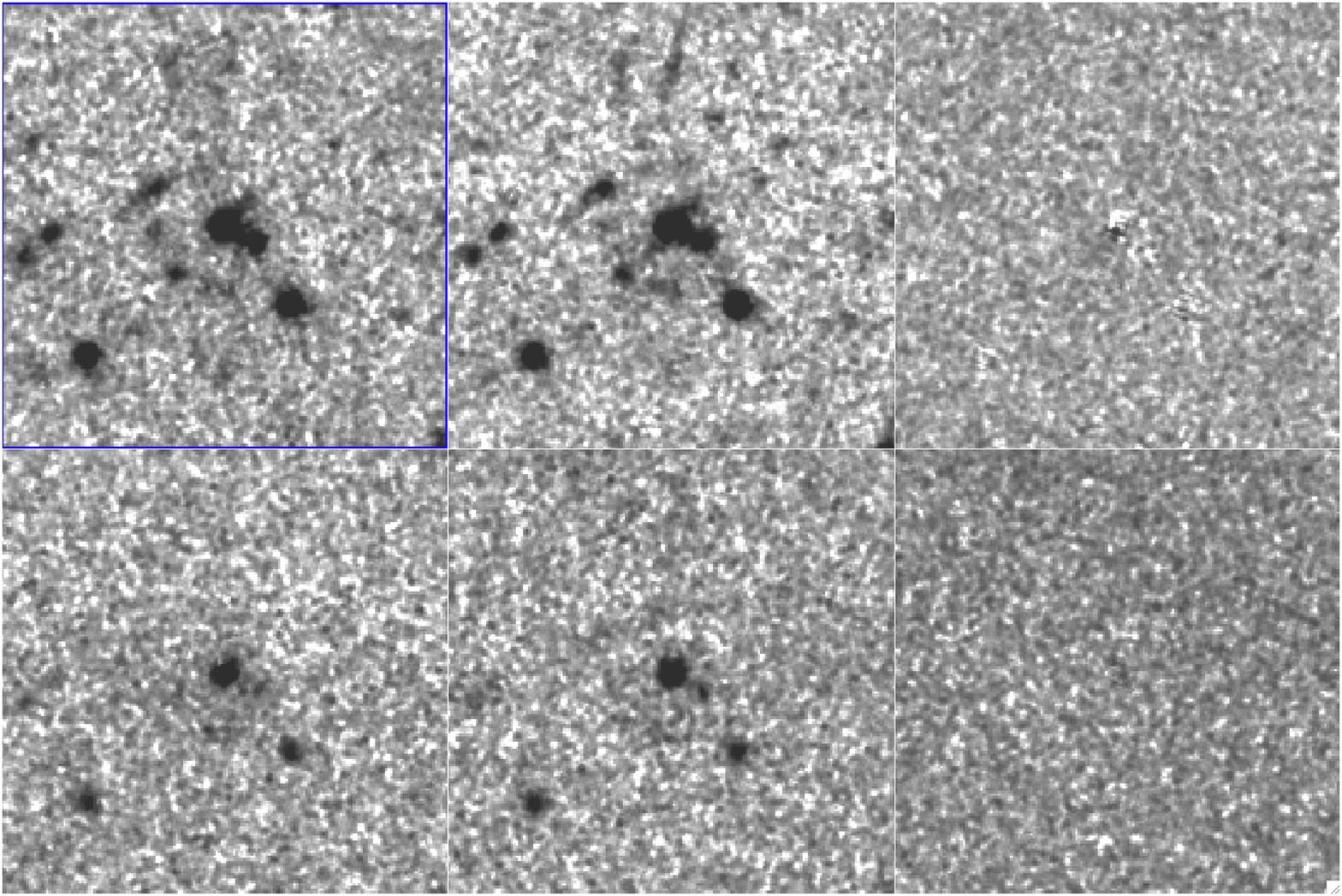}}
   \end{minipage}
\caption{(top panels) Montage of W1 images showing a 180x180 arcsec region centered on the position of 0145+234 from the AllWISE Catalog.  The left panel shows the coaddition of the January 2014 single-exposures, the center panel is the coaddition of the January 2018 single-exposures, and the right panel is the difference between the January 2018 and January 2014 images.  (bottom panels) The same is above, but for W2.
\label{fig:S2}}
\end{figure*}

\begin{figure*}
\figurenum{S3}
   \centering
   \begin{minipage}{0.99\textwidth}
   \centering{\includegraphics[width=1.0\textwidth]{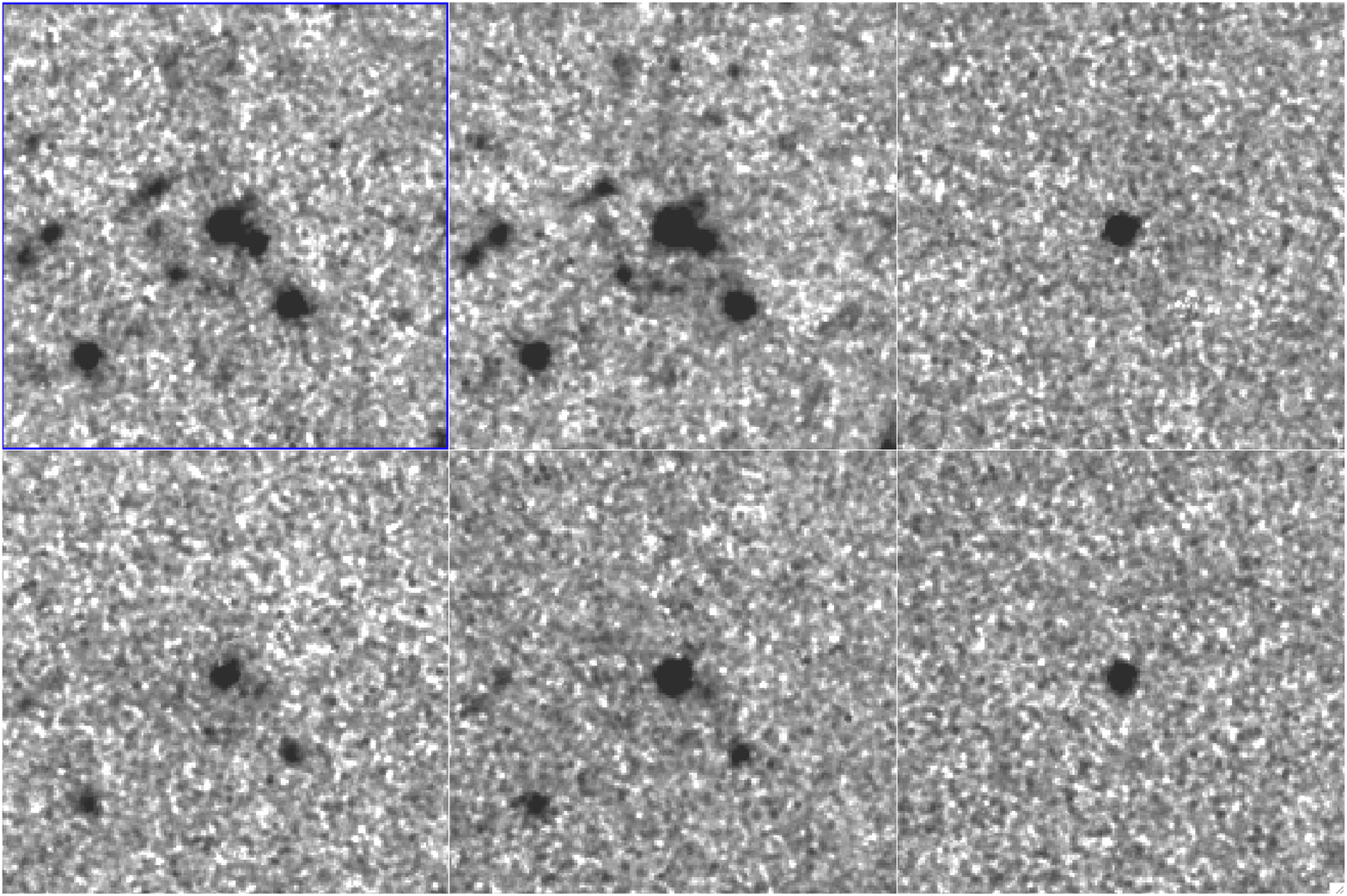}}
   \end{minipage}
\caption{(top panels) Montage of W1 images showing a 180x180 arcsec region centered on the position of 0145+234 from the AllWISE Catalog.  The left panel shows the coaddition of the January 2014 single-exposures, the center panel is the coaddition of the January 2019 single-exposures when the source was in outburst, and the right panel is the difference between the January 2019 and January 2014 images.  (bottom panels) The same is above, but for W2.
\label{fig:S3}}
\end{figure*}

\end{document}